%% file: main.tex
\documentclass[twocolumn]{./jpsj3}

\usepackage{graphicx} % for arXiv submission
\usepackage{bm}
\usepackage{braket}
\usepackage{mathrsfs}
\usepackage{comment}
\usepackage{tabularx}

\begin{document}
    \input{title.tex}

    \input{abstract.tex}
    \input{introduction.tex}
    \input{method.tex}
    \input{./result_and_discussion.tex}
    \input{./summary.tex}
    \input{./acknowledgements.tex}
    \input{./appendix.tex}
    \bibliographystyle{jpsj}
    \bibliography{main}
\end{document}

%% file: title.tex
\title{\textit{Ab initio} calculations of longitudinal electrical conductivity using 
a Wannier-based coherent potential approximation}
\date{\today}

\author{Shota Namerikawa$^1$ and Takashi Koretsune$^1$}
\inst{$^1$Department of Physics, Tohoku University, Sendai 980-8578, Japan}

%% file: abstract.tex
%\begin{abstract}
\abst{
We present a longitudinal electrical conductivity calculation method for disordered alloys applicable from a wide range of density functional theory (DFT) codes based on the first-principles Wannier-based coherent potential approximation (Wannier-CPA).
For evaluation of electrical conductivity, we employ two complementary methods; the Kubo-Greenwood formula and numerical analytic continuation of the current-current correlation function.
We apply the developed method to Ag-Pd alloys and find that the results obtained by the Wannier-CPA reasonably reproduce previous studies by the well-established CPA implementation based on the Korringa-Kohn-Rostoker Green's function method (KKR-CPA).}
    \maketitle 
%\end{abstract}

%% file: introduction.tex
\section{Introduction}
Alloying has been used to improve many physical properties of materials, including mechanical, magnetic, and transport properties.
As for thermoelectric properties, for example, alloying contributes to a high figure of merit ($zT$) by reducing phonon lattice thermal conductivity \cite{P_G_Klemens_1955, Callaway1960, Wang2013, Steigmeier1964} or 
by changing band structures to enhance the power factor \cite{Pei2012, Tan2014, Heremans2012, Tan2015, KIM2017, Liu2012}. %Giant Magnetic resistance の話も入れる？

To utilize alloying for improving properties of materials, it is important to compute the electronic structures of alloys.
The supercell method \cite{Zunger1990} is a naive approach, where substitution is modeled by considering
a periodically repeated large unit cell. However, the computational cost of this method becomes expensive with the size of supercells and the artifacts of periodicity should be carefully treated.
On the other hand, there are some methods that can evaluate the electronic structures of alloys while keeping their unit cell size fixed, such as
the virtual crystal approximation (VCA)\ \cite{Neddermeyer1984} and the coherent potential approximation (CPA)\ \cite{Soven1967}.  
The VCA simply mixes the potentials of substituted atoms and is frequently used in evaluations of band structures of alloys. However, this approach is known to produce a physically wrong density of states (DOS) in model calculations \cite{Velicky1968}.
By contrast, the CPA can generate more natural behavior in the DOS of alloys than the VCA \cite{Velicky1968}.
Furthermore, the CPA can include the effect of a lifetime due to disorder scattering in the self-energy of Green's functions, which cannot be incorporated by the VCA. 

The CPA has been utilized for calculations of transport properties in alloys.
Velicky first introduced an approach to calculate conductivity in the CPA scheme and applied the method to empirical two-level
tight-binding models \cite{Velicky1969}. Furthermore, Butler developed a method similar to Velicky's, but more specialized in the KKR-CPA \ \cite{Butler1985} and Turek proposed a conductivity calculation method that combines the tight-binding linearized muffin-tin orbital (TB-LMTO) approach with the CPA.
These two methods are commonly employed for transport calculations based on the CPA combined with DFT\ \cite{Swihart1986, Fukushima2022}. 
Although these methods are initially formulated using the Kubo-Greenwood formula, which is applicable only to the symmetric components of the conductivity tensor\ \cite{Ebert2011}, they have been extended to the Kubo-Bastin or the Kubo-Streda formulas. 
By using these two formulas, more general parts of conductivity tensors in alloys have been computed including parts important in spintronics such as anomalous Hall conductivity\ \cite{Lowitzer2010, Turek2014, Hyodo2016} and spin Hall conductivity\ \cite{Lowitzer2011, Turek2019}.
In addition to these extensions to transverse conductivity, transport calculations at finite temperatures have progressed by employing the alloy analogy model in the CPA framework\ \cite{Ebert2015}.  
This model is exploited to assess the scattering effect on conductivity at finite temperatures derived from thermal fluctuation of spin direction and atomic displacements due to thermal vibration\ \cite{Drchal2018, Sakuma2018, Sakuma2022, Shinya2023, Nam2024}.
Though these techniques broaden the possibility of transport calculations in alloys, they are not easily available from DFT methods other than the KKR or the TB-LMTO methods. 

Recently, a first-principles Wannier-based CPA method (Wannier-CPA) was developed \cite{Ito2022}. 
The Wannier-CPA can be executed from any DFT codes that can export the information to construct Wannier functions, especially maximally-localized Wannier functions (MLWF).
This method has been confirmed to reproduce physical quantities such as Bloch spectral functions, density of states, and magnetization in iron-based transition metal alloys.
Since the MLWF technique is used for characterizing the systems from first principles and serves as a starting point of many post-processing techniques\ \cite{PIZZI2014422, Qiao2018, Ryoo2019, Tsirkin2021}, extending the Wannier-CPA method to transport calculations can broaden the utility of the MLWF approach.
%However, the transport calculations of alloys have not yet been executed based on the CPA using MLWFs, although MLWFs are 
%widely used as a computationally effective tool to calculate numerous physical properties including transport properties of perfect crystals \cite{PIZZI2014422, Qiao2018, Ryoo2019, Tsirkin2021}.
% Wannierでできると何がいいかをもう少しうまく主張したい 

In this study, we develop a method to obtain the electrical conductivity of disordered substitutional alloys 
based on the Wannier-CPA as a first step towards extending the availability of transport calculations in alloys from DFT codes. 
We use two conductivity calculation methods; the Kubo-Greenwood formula and numerical analytic continuation of the current-current correlation function on the Matsubara axis. 
Although the latter method employs numerical analytic continuation, which is known to be an ill-conditioned problem\cite{Zhen2023}, it offers several advantages. 
These include the capability to treat correlation effects using techniques such as the dynamical mean field theory\cite{Minar2005}, and enhanced computational efficiency by performing CPA calculations at complex energy points\cite{Hass1984}.
We apply the developed method to Ag-Pd alloys and evaluate their resistivity to confirm that 
our method is practical for electrical conductivity calculations in alloys. 

%% file: method.tex
%calculation flow
\section{Method} \label{seq:Method}
In this section, we explain how to compute the electrical conductivity of alloys.
The calculation flow is as follows. First, we obtain the coherent Green's functions of alloys by employing the Wannier-CPA method. Next, we evaluate the conductivity using the Kubo-Greenwood formula and numerical analytic continuation of the current-current correlation function.
We omit band and orbital indices in the equations below for simplicity.

\subsection{Wannier-based CPA}\label{seq:Wannier-CPA}

First, we briefly review how to derive the coherent Green's functions in the Wannier-CPA. 
Although, we restrict our explanation to the binary-alloy ($A_x B_{1-x}$) cases with one atom in a unit cell for brevity, extension to multi-component cases is straightforward.

In the CPA, we determine the coherent potential, $\sigma_c (z)$ that satisfies the equation for impurity averaged single-site scattering matrix,
\begin{equation}
    \braket{t(z)}_{\mathrm{imp}} \equiv x t^A(z) + (1-x) t^B(z) = 0,
    \label{eq:CPA condition}
\end{equation}
where
\begin{equation}
    \begin{split}
    &t^{A (B)}(z) = \\
    &(\varepsilon_{0}^{A (B)} - \sigma_c(z)) \left[\bm{1} - G_c(z) (\varepsilon_{0}^{A (B)} - \sigma_c(z))\right]^{-1},
    \end{split}
    \label{eq:t-matrix definition}
\end{equation}
is the single-site scattering matrix of $A(B)$ atom, $z$ is energy in complex plane, $\bm{1}$ is the identity matrix, and $\varepsilon_0^{A(B)}$ is an onsite energy of $A(B)$ atom. The coherent Green's function is defined as
\begin{equation}
    G_c(z) = \frac{1}{\Omega_\mathrm{BZ}}\int{d\bm{k} G_{c}(\bm{k},z)},
\end{equation}
with
\begin{equation}
    G_c (\bm{k}, z) = \frac{1}{ (z + \mu)\bm{1} - \sigma_c(z) - H(\bm{k})},
    \label{eq:coherent Green function}
\end{equation}
where $\Omega_\mathrm{BZ}$ is the volume of the Brillouin zone, $\mu$ is the chemical potential, and
\begin{equation}
    H(\bm{k}) = \sum_{\bm{R}\neq \bm{0}}{e^{i\bm{k}\cdot \bm{R}} H(\bm{R})},
    \label{eq:hopping term}
\end{equation}
is the Fourier transformation of the hopping terms in the tight-binding Hamiltonian.
In the calculations of the hopping terms $H(\bm{R}\neq 0)$ in Eq.~\eqref{eq:hopping term}, we take the configuration average as
\begin{equation}
    H(\bm{R}) = x H^{A}(\bm{R}) + (1-x) H^{B}(\bm{R}),
    \label{eq:hopping average}
\end{equation}
where $H^{A(B)}(\bm{R})$ are the hopping terms in the pure crystals composed of A(B) atoms for simplifying the dependency of Green's functions on atomic configurations.
We calculate $\sigma_c (z)$ that satisfies Eq.~\mbox{\eqref{eq:CPA condition}} by following the algorithm used in Ref. \citen{Ito2022}. Once $\sigma_c (z)$ is obtained, the
coherent Green's function can be determined from Eq.~\mbox{\eqref{eq:coherent Green function}}. 

\subsection{Electrical conductivity calculations}\label{sec:electrical_cond}
After obtaining the coherent Green's functions, we calculate electrical conductivity from these Green's functions.
Diagonal components of electrical conductivity can be evaluated from the correlation function, $\Phi_{ii}(\hbar \omega + i \delta)$, as 
\begin{equation}
    \sigma_{ii} = \lim_{\omega \rightarrow 0} \frac{ \Phi_{ii} (\hbar \omega + i\delta) - \Phi_{ii} (0)}{i\omega},
    \label{eq:relation between sigma and correlation}
\end{equation}
\begin{equation}
    \Phi_{ii}(\hbar \omega + i\delta) = \Phi_{ii}(i\omega_\nu)\left. \right|_{(i\omega_\nu \rightarrow \hbar \omega + i \delta)},
    \label{eq:real freq Phi}
\end{equation}
where $\hbar$ is the Dirac constant and $\delta$ is an infinitesimal positive constant, respectively. 
When neglecting the vertex corrections originating from the configuration average\ \cite{Velicky1969,Butler1985}, $\Phi_{ii} (i\omega_\nu)$ is given as \cite{MahanTextbook}
\begin{equation}
    \begin{split}
    &\ \Phi_{ii} (i\omega_\nu) \\
    &= - \frac{k_\mathrm{B} T e^2}{V}\sum_{\bm{k},n}{\mathrm{tr}\left[v^i_{\bm{k}} G_c(\bm{k},i\varepsilon_n - i \omega_\nu) v^i_{\bm{k}} G_c (\bm{k}, i\varepsilon_n)\right]}, \\
    &
    \end{split}
    \label{eq:c-c correlation}
\end{equation}  
where $k_\mathrm{B}$ is the Boltzmann constant, $T$ is absolute temperature, $V$ is the volume of the system, $e$ is the elementary charge, $i \omega_\nu$ and $i\epsilon_n$ are bosonic and fermionic Matsubara frequencies, respectively.
For the velocity operators, $v^i_{\bm{k}}$, we take the configuration average like Eq.\eqref{eq:hopping average} as 
\begin{equation}
    \begin{split}
    v^i_{\bm{k}} &\equiv \frac{1}{\hbar}\frac{\partial H(\bm{k})}{\partial \bm{k}}\\ 
                 &= \frac{1}{\hbar}\sum_{\bm{R}}{i R_i e^{i\bm{k}\cdot \bm{R}} (x H^{(A)} (\bm{R}) + (1-x) H^{(B)} (\bm{R}))}. \\
                 &
    \end{split}
\end{equation}  
Note that the velocity operators and the coherent Green's functions in Eq.~\mbox{\eqref{eq:c-c correlation}} are evaluated in the Wannier-gauge, while the result is gauge-independent.
We ignore the inter-band matrix elements of the current operator in the Hamiltonian gauge since we focus on the longitudinal conductivity, where the inter-band terms do not give significant contributions. 

%computational implementation
To evaluate electrical conductivity, we use two approaches: the Kubo-Greenwood formula and numerical analytic continuation of the current-current correlation function in this study.
The Kubo-Greenwood formula is the analytic expression of Eq. \eqref{eq:relation between sigma and correlation} at $T = 0\ \mathrm{K}$ and is given as \cite{Velicky1969,Butler1985} 
\begin{equation}
    \begin{split}
        & \sigma_{ii}(T = 0\ \mathrm{K}) \\ 
        = &-\frac{e^2\hbar}{4\pi V} \sum_{\bm{k}}{\mathrm{tr}[v^i_{\bm{k}} (G^{(R)}_c(\bm{k},E_\mathrm{F}) - G^{(A)}_c(\bm{k},E_\mathrm{F}))}  \\
        & \quad \quad \quad \quad \quad \quad\ v^i_{\bm{k}} (G^{(R)}_c(\bm{k},E_\mathrm{F}) - G^{(A)}_c(\bm{k},E_\mathrm{F}))], %VelickyとButlerを引用する。
    \end{split}
    \label{eq:cpa conductivity at T=0K}
\end{equation}
where 
\begin{equation}
    G^{(R/A)}_c(\bm{k}, \varepsilon) = \frac{1}{(\varepsilon \pm i\delta)\bm{1} - \sigma_c(\varepsilon \pm i\delta) - H^c(\bm{k})},
    \label{eq:Green's function on real axis}
\end{equation}
and $E_\mathrm{F}$ is the Fermi energy. Note that the vertex corrections are neglected in Eq.~\mbox{\eqref{eq:cpa conductivity at T=0K}}.

On the other hand, in the current-current correlation function approach, we first calculate Eq.~\mbox{\eqref{eq:c-c correlation}} numerically.
To sum over the Matsubara frequencies in Eq.~\mbox{\eqref{eq:c-c correlation}}, we use the intermediate representation and the sparse sampling method \cite{Shinaoka2017,Chikano2018,Li2020}.
In this approach, we first consider the imaginary-time correlation function defined as
\begin{equation}
\Phi_{ii} (\tau) = - \frac{e^2}{V}\sum_{\bm{k}}\mathrm{tr}[v^i_{\bm{k}} G_c(\bm{k},\tau) v^i_{\bm{k}} G_c (\bm{k},-\tau)],
\label{eq:phi in tau}
\end{equation}
where $G_c (\bm{k},\tau)=\frac{1}{\beta}\sum_{n}{e^{-i\varepsilon_n \tau} G_c (\bm{k},i\varepsilon_n)}$.
Next, we calculate $\Phi_{ii} (\tau)$ in Eq.~\mbox{\eqref{eq:phi in tau}} and convert the function efficiently into the correlation function at sparse-Matsubara frequencies, $\Phi_{ii}(i\omega_\nu)$ in Eq.~\mbox{\eqref{eq:c-c correlation}} by employing the SparseIR.jl package\ \cite{Shinaoka2017,Li2020,Wallerberger2023}.
After obtaining $\Phi_{ii}(i\omega_\nu)$, we execute numerical analytic continuation of $\Phi_{ii}(i\omega_\nu)$ to calculate electrical conductivity from Eq.~\mbox{\eqref{eq:relation between sigma and correlation}}. In this process, since tiny noise in $\Phi_{ii}(i\omega_\nu)$ can affect the results, 
several methods should be used to guarantee the correctness of the analytic continuation. 
In this study, we use two numerical analytic continuation methods: Pad\`{e} approximation and Nevanlinna analytic continuation\ \cite{Fei2021}.
The latter method has recently proven to be applicable to bosonic functions such as the longitudinal current-current correlation function by converting the functions into auxiliary fermionic ones \cite{Nogaki2023}.

\subsection{Computational conditions}
In this research, we calculate the longitudinal resistivity\ ($\rho_{xx} = \sigma_{xx}^{-1}$) in Ag-Pd alloys as a model case
to verify effectiveness of the Wannier-CPA method in electrical conductivity calculations.
We choose this material because there are several studies evaluating the resistivity of the alloys using the KKR-CPA \cite{Butler1984,Banhart1998},
the TB-LMTO-CPA \cite{Turek2002} and the experiment\ \cite{Gueanault1974}.
Perfect crystals of both elements have face-centered cubic crystal structures. The lattice constant of Ag(Pd) is set as 4.253 \AA(3.951 \AA). %実験のreferenceが必要？
We use Quantum\textsc{Espresso} \cite{QUANTUM_ESPRESSO} for DFT calculations.
We employ the ultrasoft pseudopotentials\ \cite{Vanderbilt1990} with Perdew-Burke-Ernzerhof exchange-correlation functional\ \cite{PBE1996} for band calculations of simple-substance Ag and Pd. 
Furthermore, we construct the tight-binding Hamiltonians of both elements using Wannier90 \cite{Wannier90} from the results of the DFT calculations.
As the initial projections for the Wannierization process, we select $s,p,d$ orbitals corresponding to configurations of $5s, 5p, 4d$ orbitals in pseudopotentials of both Ag and Pd.
We obtain one-shot Wannier functions; that is, we only execute disentanglement process to preserve orbital-like characters of the Wannier functions.
Since the DFT-based Wannier Hamiltonian does not contain the information of the reference energy of the model, we have to determine the difference of onsite energies of Ag and Pd used in CPA calculations\ \cite{Ito2022}.
We determine the value by calculating the difference in onsite energy for each atom, measured relative to the electrostatic potential in the vacuum layers obtained from slab calculations for Ag and Pd, as detailed in Appendix~\ref{appendix: onsite}.

In resistivity evaluations, we set the volume of the alloys in Eqs.\eqref{eq:c-c correlation} and \eqref{eq:cpa conductivity at T=0K} as concentration average of Ag and Pd.
For numerical analytic continuation, we use the sparse frequency sampling points generated by the SparseIR package  \cite{Shinaoka2017,Li2020,Wallerberger2023}.
The number of Matsubara frequency points used for analytic continuation was determined by carefully examining the dependence of the results on the number of points, as shown in \mbox{Appendix \ref{sec:AC vs sample}}. 
Consequently, we employ 30 sampling points near the origin for the Padé approximation and 10 points for the Nevanlinna analytic continuation.
The analytic continuation is executed at 256-digit precision to obtain precise results.
For the calculation based on the Kubo-Greenwood formula, we set small imaginary part, $\delta$ in Eq.~\mbox{\eqref{eq:Green's function on real axis}}, to $10^{-3}\ \mathrm{eV}$.
To obtain converged value of electrical conductivity, we use around $100^3$ $\bm{k}$  points for calculations of the current-current correlation functions in Eq.~\mbox{\eqref{eq:c-c correlation}}, whereas we use around $200^3$ $\bm{k}$  points for calculations of electrical conductivity using the Kubo-Greenwood formula in Eq.~\mbox{\eqref{eq:cpa conductivity at T=0K}}.
The dependence of resistivity value on the number of $\bm{k}$  points calculated by using the two methods is shown in Appendix \ref{sec: convergence w.r.t. k  points}.
%It is worth mentioning that computational cost per k-point can be reduced because of Wannier formalism in CPA.

%% file: result_and_discussion.tex
\section{Results and Discussion}
\label{sec: Results and Discussion}
%最初の部分はmethodの最後にcomputational detailとして回す。
\subsection{Electronic structures of Ag and Pd}
First, we show the band structures of Ag and Pd obtained by DFT and their Wannier interpolation.
\begin{figure}
    \includegraphics[width=\linewidth]{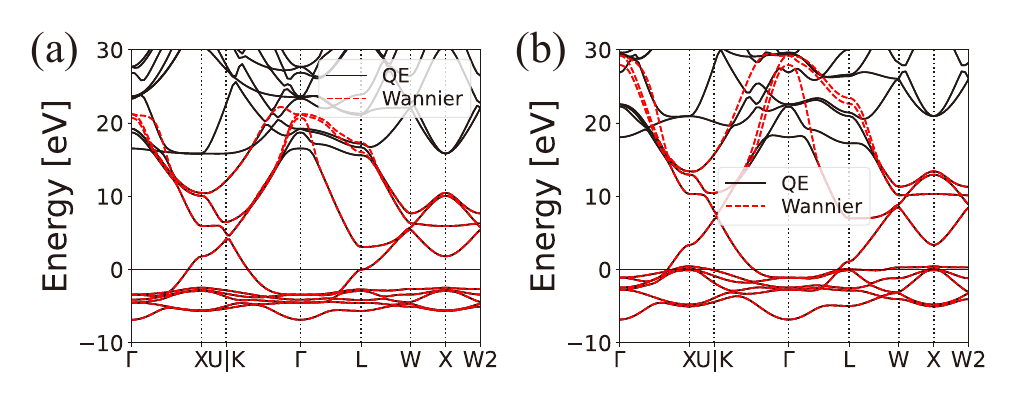}
    \caption{Band structures of (a) Ag and (b) Pd. The black lines represent the DFT band structures, the red lines indicate the Wannier-interpolated band structures, and the blue line marks the Fermi energy.}
    \label{fig: Ag and Pd bands}
\end{figure}
In Fig.~\mbox{\ref{fig: Ag and Pd bands}}, the Fermi energy in Ag is located within $s$-orbital-like energy bands, while the Fermi energy in Pd is located on the upper bound of $d$-orbital-like energy bands with low dispersion in addition to $s$-orbital like bands.
Since incoherent scattering occurs when energy of $s$-orbital bands and that of $d$-orbital bands overlap around the Fermi energy\ \cite{Banhart1998},
resistivity is expected to be relatively low in the Pd-rich region and relatively high in the Ag-rich region.
Note that we set the upper bound of the inner window in the Wannierization as high as possible (15 eV above the Fermi energy in Ag and Pd) while keeping the spreads of Wannier functions reasonably small. 
This is because the band energies both below and above the Fermi energy contribute to the onsite energies of Ag and Pd. Since the onsite energy differences between Ag and Pd are crucial for CPA calculations, 
the precise determination of the onsite energies with an appropriate inner window is important for this method.
\subsection{Comparison of the Kubo-Greenwood formula and numerical analytic continuation of the correlation function}
\begin{figure}[tbp]
    \centering
    \includegraphics[width=\linewidth]{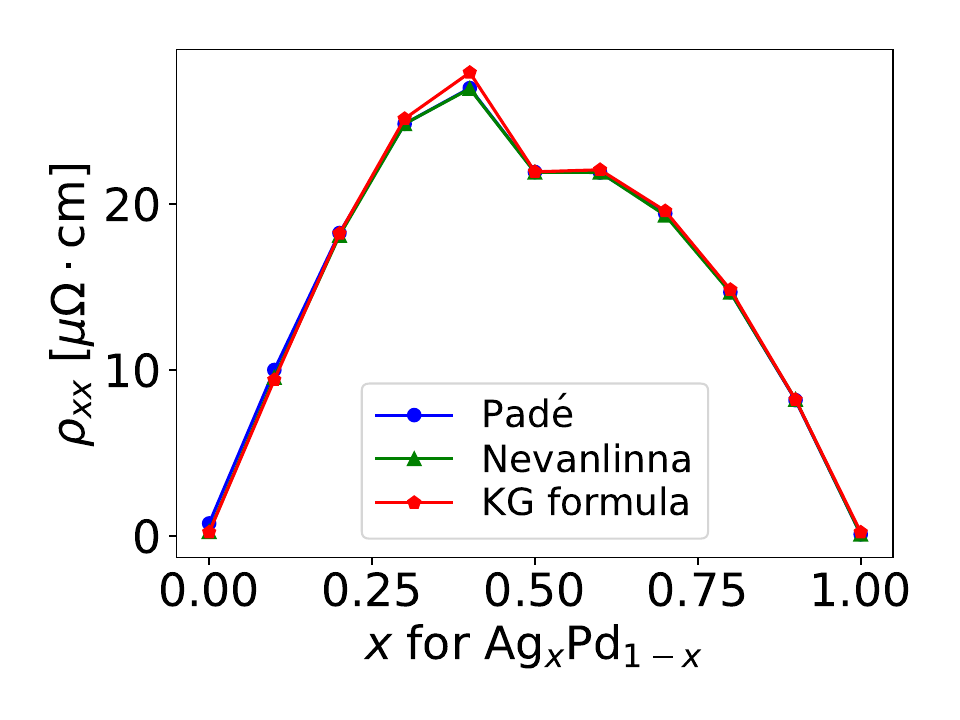}
    \caption{
            Resistivity of $\mathrm{Ag}_x \mathrm{Pd}_{1-x}$ alloys as a function of concentration $x$ calculated using
            the Kubo-Greenwood formula (red line), analytic continuation at $T = 300\ \mathrm{K}$ with Pad\'{e} approximation (blue line) and Nevanlinna analytic continuation (green line).  
             }
    \label{fig:comparison_of_AC_Ag-Pd}
\end{figure}

Figure \ref{fig:comparison_of_AC_Ag-Pd} shows a comparison of the resistivity of Ag$_x$ Pd$_{1-x}$ alloys calculated using the Kubo-Greenwood formula($T = 0\ \mathrm{K}$) and the two numerical analytic continuation methods($T = 300\ \mathrm{K}$): Pad\'e approximation and Nevanlinna analytic continuation. 
Clearly, the two analytic continuation methods yield nearly identical results, indicating that the numerical analytic continuation works quite well. 
The results obtained from the Kubo-Greenwood formula are slightly higher than those of analytic continuation methods particularly at $x = 0.4$.

\begin{figure}[tbp]
    \centering
    \includegraphics[width=\linewidth]{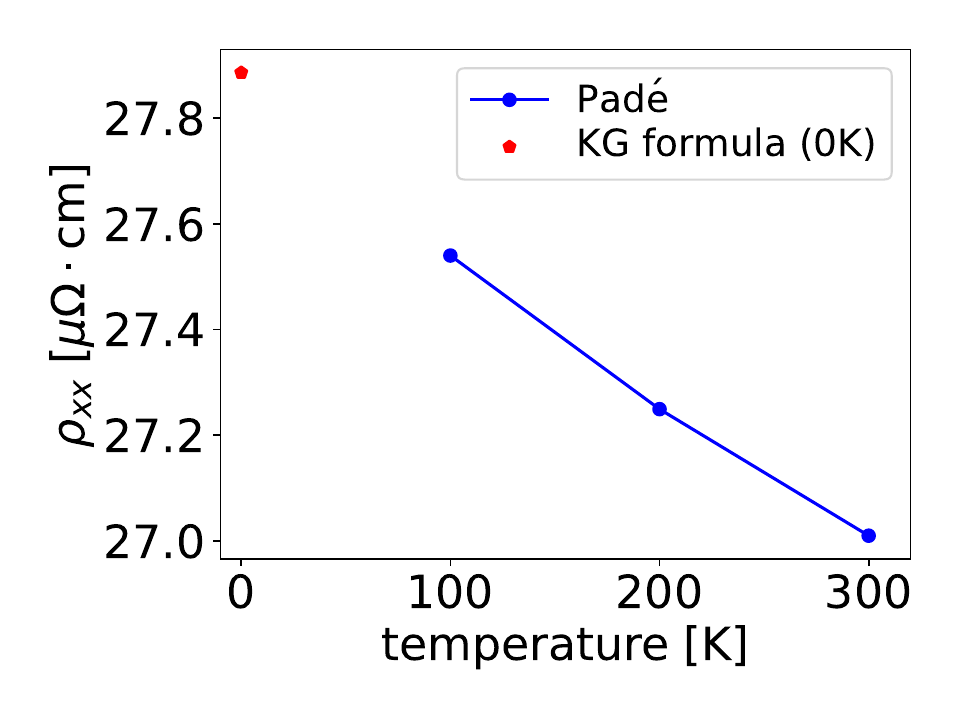}
    \caption{
            Resistivity of $\mathrm{Ag}_{0.4} \mathrm{Pd}_{0.6}$ alloys as a function of temperature calculated by employing Pad\'{e} approximation\ (blue line).
            Resistivity at $T = 0\ \mathrm{K}$ evaluated using the Kubo-Greenwood formula is also shown($\rho_{xx} = 27.89\ \mu \mathrm{\Omega \cdot cm}$) for comparison (red point). 
            The number of $\bm{k}$ points for CPA and conductivity calculations is $200^3$.
    }
    \label{fig:temperature effect}
\end{figure}

To understand this discrepancy, we evaluate resistivity at different temperatures using analytic continuation methods. 
In Fig.~\mbox{\ref{fig:temperature effect}}, we show the resistivity of Ag$_{0.4}$Pd$_{0.6}$ as a function of temperature, evaluated using Pad\`e approximation.  
As seen from Fig.~\mbox{\ref{fig:temperature effect}}, the resistivity increases as temperature decreases and approaches the value calculated by using the Kubo-Greenwood formula 
at $T = 0\ \mathrm{K}$. This suggests that the deviation between the Kubo-Greenwood formula and the analytic continuation in Fig.~\mbox{\ref{fig:comparison_of_AC_Ag-Pd}} is primarily due to the temperature.

\subsection{Comparison with previous studies}
\begin{figure}[tbp]
    \centering
    \includegraphics[width=\linewidth]{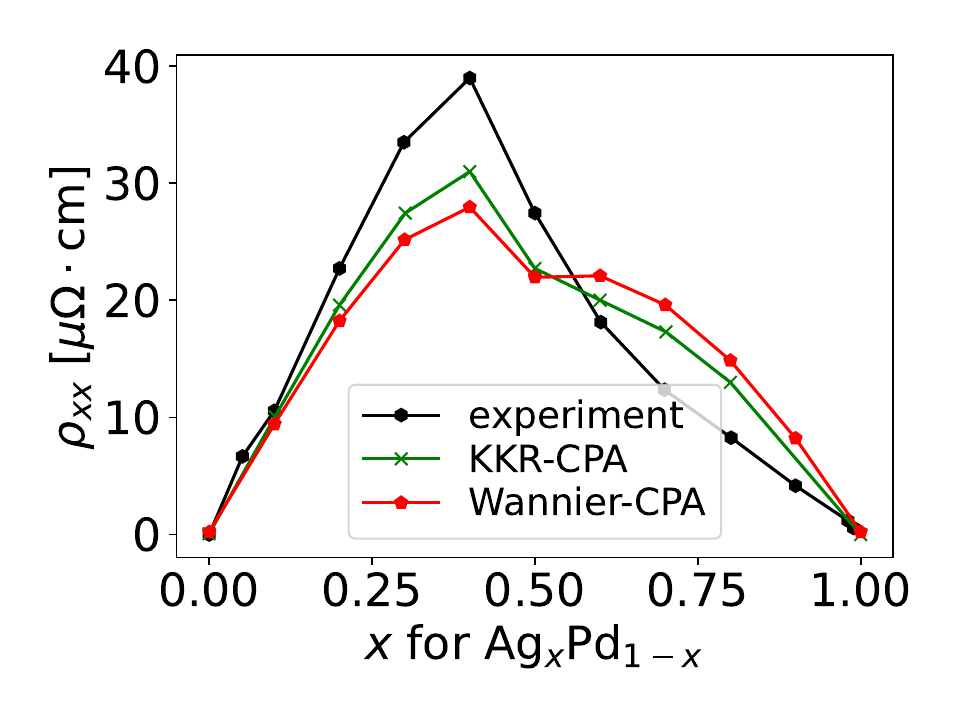}                    
    \caption{
        Resistivity of $\mathrm{Ag}_x \mathrm{Pd}_{1-x}$ alloys as a function of concentration $x$ evaluated using the Kubo-Greenwood formula in this study (red line), and by  
        the KKR-CPA method with the Kubo-Greenwood formula (green line)\cite{Tulip2008}.
        The experimental value at $T = 4.2\ \mathrm{K}$ (black line) is also shown\cite{Gueanault1974}.
        }
    \label{fig:compare_with_previous_study_Ag-Pd_rho}
\end{figure}

Next, we present a comparison of the results of the resistivity calculations using the Wannier-CPA method with Pad\'e approximation, the results using the KKR-CPA method\ \cite{Tulip2008}, and an experiment\ \cite{Gueanault1974} in Fig.\ \ref{fig:compare_with_previous_study_Ag-Pd_rho}.
As seen from Fig.\ \ref{fig:compare_with_previous_study_Ag-Pd_rho}, the results generated by the Wannier-CPA reproduce the characteristic features of the concentration dependence of the resistivity in 
KKR-CPA calculations without the vertex corrections and in the experiment such as the peak position at $x=0.4$.
The maximum relative difference between the Wannier-CPA method and the KKR-CPA method for each concentration of Ag is approximately 14\%, which can be considered a good agreement given the differences between the two methodologies.
%Soven(1978)を引用？
%Therefore, we concluded Wannier-CPA has comparable precision for electrical-conductivity calculation with that of the KKR-CPA
%for calculations in which vertex correction is neglected. 
%The result well reproduces the character of resistivity dependence on concentration
%like peak position and shoulder structure at $x = 0.5$. 
%We checked analytic continuation works well by comparing Pade approximation and Nevanlinna analytic continuation. 
%By considering difference of formalism between the KKR-CPA and Wannier-CPA, deviation of different methods is small.  

%% file: summary.tex
\section{Conclusion}
In this study, we have developed an electrical conductivity calculation method within the Wannier-CPA formalism.
To confirm the precision of the method, we evaluated the resistivity of Ag-Pd alloys.
We found that our results closely matched those obtained using the KKR-CPA method. This indicates that the Wannier-CPA approach is effective for longitudinal electrical conductivity calculations of alloys. 
Our method is available from any DFT codes that can generate Wannier functions, and can be easily combined with any model Hamiltonian techniques using either the imaginary time formalism or real frequency formalism.
Thus, we believe that this method will broaden the accessibility and functionality of conductivity calculations for alloys, which will be useful for investigating transport phenomena of alloys, especially using high-throughput calculations\ \cite{Anubhav2013,PIZZI2016,Sakai2020}.

%% file: acknowledgements.tex
\begin{acknowledgments} 
    The authors thank K. Nogaki, H. Shinaoka, and F. Kakizawa for their fruitful discussions. This work was supported by JST SPRING, Grant Number JPMJSP2114,
    JSPS KAKENHI Grant No. 21H01003, 21H04437, 22K03447, and 23H04869, JST-Mirai Program (JPMJMI20A1), JST-ASPIRE (Grant No. JPMJAP2317),
    Center for Science and Innovation in Spintronics (CSIS), Tohoku University.
    S.N. acknowledges support from GP-MS at Tohoku University.
\end{acknowledgments}

%% file: appendix.tex
\appendix
\section{Determination of Onsite-energy Difference Using Slab Calculations}
\label{appendix: onsite}
\begin{figure}[tbp]
\centering
\includegraphics[width=\linewidth]{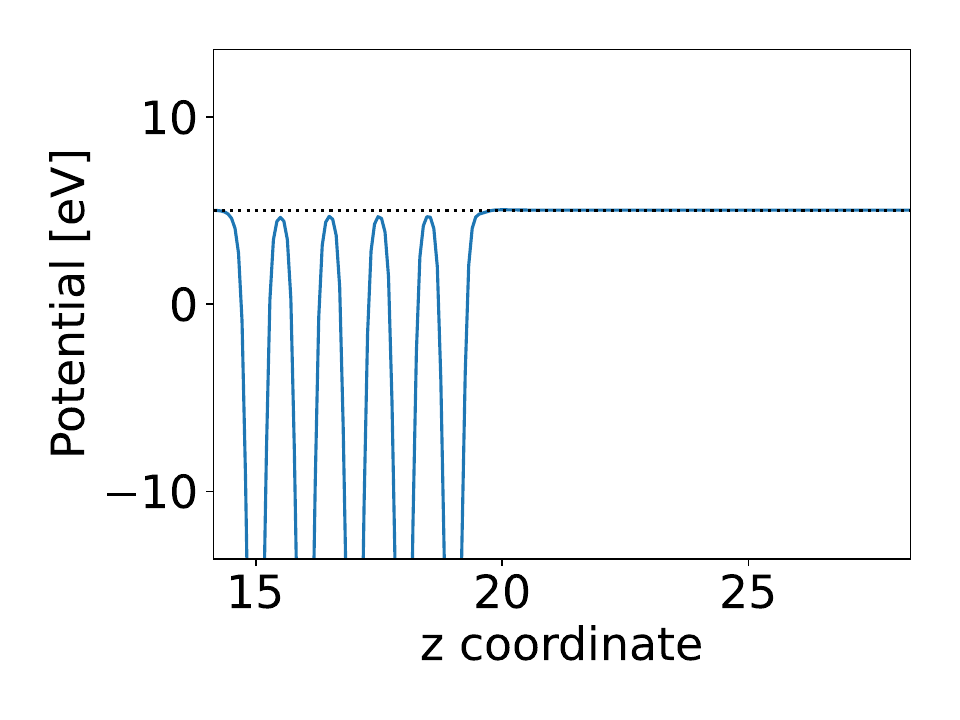}
\caption{
    Plane-averaged electrostatic potential as a function of the z coordinate perpendicular to the (001) surface 
    in units of the lattice constant of Ag. 
    The flat-energy region corresponds to the vacuum layer and the black
    horizontal dotted line represents the vacuum level.
    }
\label{fig: slab calculation}
\end{figure}

We describe the method for determining the onsite energy difference of Wannier functions between Ag and Pd. 
First, we perform slab calculations for the pure crystals of Ag and Pd. 
The slab is constructed by repeating a bulk crystal layer five times along the direction perpendicular to (001) surface
and adding a vacuum layer approximately twice as thick as the slab layers. 
For each slab, we calculate the value of the plane-averaged electrostatic potential 
in the vacuum region as shown in Fig. \ref{fig: slab calculation}, and use the potential value as the reference vacuum energy for each atom. \cite{Guosheng2021, Ying2017, Kondo2019}
After obtaining the reference energy, we align these values between Ag and Pd and calculate 
the difference in the average of onsite energy of the $d$-orbital-like Wannier functions 
at the innermost part of the slabs. We choose $d$ orbitals due to their minimal spatial extension, 
making them less susceptible to the influence of the vacuum layer. 
Before performing CPA calculations, we adjust the bulk onsite energy difference to match the value obtained 
from the slab calculations, following the approach outlined in Ref. \citen{Ito2022}.
\section{Results of Analytic Continuation for Different Number of Sampling Points}
\label{sec:AC vs sample}
\begin{figure}[tbp]
\centering
\includegraphics[width=\linewidth]{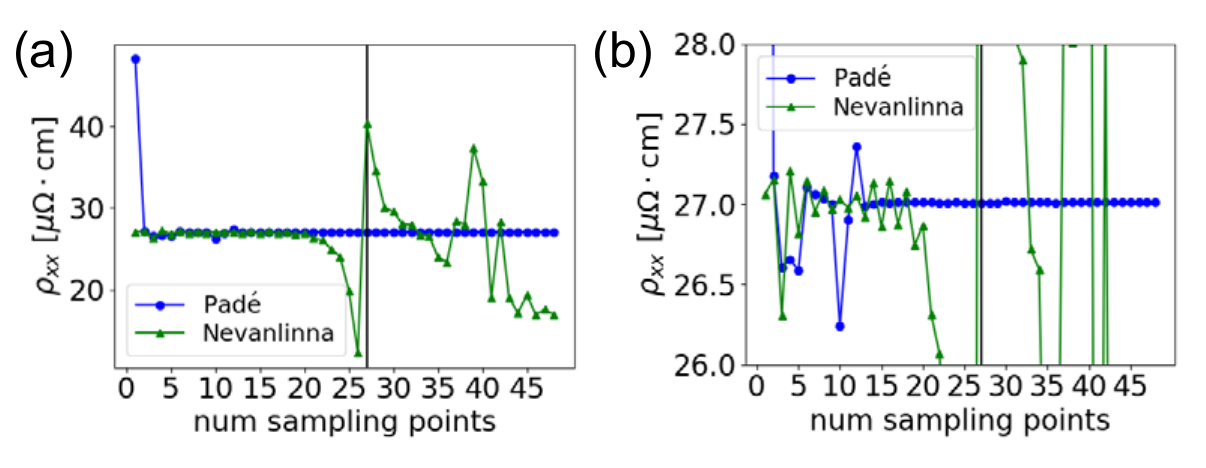}
\caption{(a) Resistivity of Ag$_{0.4}$Pd$_{0.6}$ evaluated using Pad\'e approximation (blue line) and Nevanlinna analytic continuation (green line) as a function of the number of sampling points for analytic continuation.
The vertical line at 27 sampling points represents the maximum number of sampling points that satisfies the Pick criterion\ \cite{Fei2021}.
The resistivity values for $200^3$ $\bm{k}$ points at $T = 300\ \mathrm{K}$ are shown.
(b) Same as (a) but with resistivity range from $26\ \mu \Omega \cdot \mathrm{cm}$ to $28\ \mu \Omega \cdot \mathrm{cm}$.
}
\label{fig:rho_vs_sampling points}
\end{figure}

We discuss the validity of the two analytic continuation methods; Pad\'e approximation and Nevanlinna analytic continuation, by changing the number of Matsubara frequency sampling points.
Figure \ref{fig:rho_vs_sampling points} shows the resistivity obtained by the two methods as a function of the number of sampling points.
Clearly, there is a significant difference between the two methods in their dependence on the number of sampling points.
The result of Pad\'e approximation is unstable when the number of sampling points is fewer than 13 while it becomes almost constant when the number is sufficiently large.
In contrast, the Nevanlinna analytic continuation provides relatively stable results with around 10 sampling points but starts to deviate when the number of points approaches 20.
The deviation becomes enormous when the number of sampling points is close to the maximum number that satisfies the Pick criterion\ \cite{Fei2021}. 
It is remarkable that the results of the two different analytic continuation methods converge to the almost identical value when sampling points are properly chosen, indicating that the converged value is quite reliable.

\section{Convergence of Resistivity with respect to the Number of $k$ Points}
\label{sec: convergence w.r.t. k points}
\begin{figure}[tbp]
    \centering
    \includegraphics[width=\linewidth]{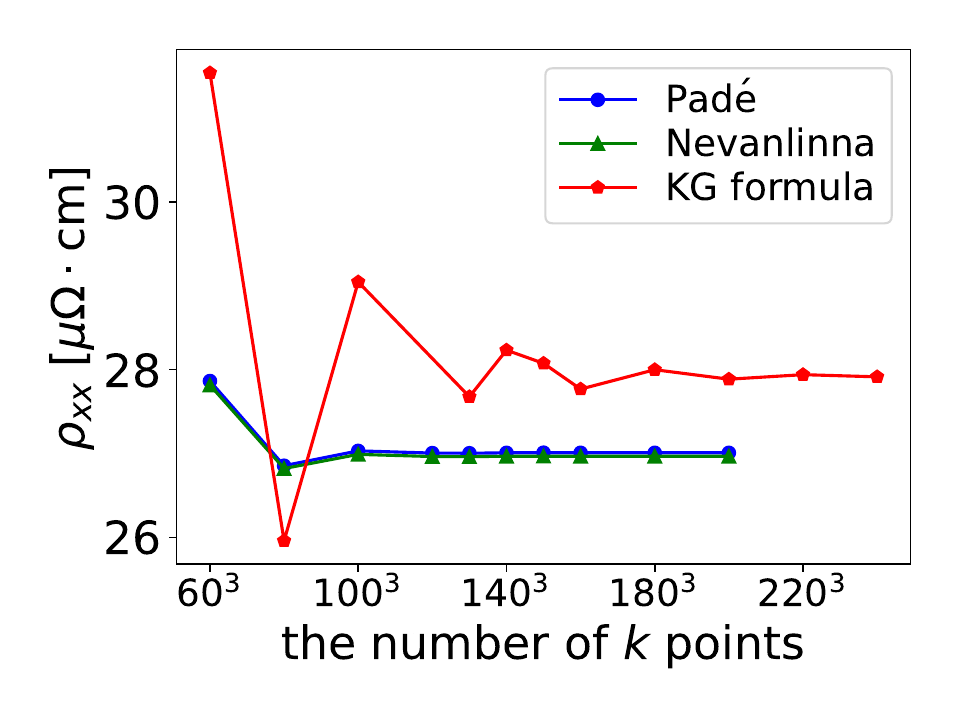}
    \caption{Resistivity of Ag$_{0.4}$Pd$_{0.6}$ as a function of the number of $\bm{k}$ points obtained by analytic continuation of the correlation function using Pad\'e approximation\ (blue line) and
    Nevanlinna analytic continuation\ (green line), and the Kubo-Greenwood formula (red line).
    }
    \label{fig:resistivity vs kpoints}
\end{figure}

Here, we discuss the convergence with respect to the number of $\bm{k}$ points of resistivity. % We discussを２回続けたくない。
Figure\ \ref{fig:resistivity vs kpoints} shows the resistivity of Ag$_{0.4}$Pd$_{0.6}$ alloys calculated using Pad\'e approximation and Nevanlinna approximation at $T=300\ \mathrm{K}$, and the Kubo-Greenwood formula with $\delta = 1\ \mathrm{meV}$.
It is observed that the resistivity evaluated using the Kubo-Greenwood formula converges around $200^3$ $\bm{k}$ points, whereas the resistivity obtained using analytic continuation methods converges more rapidly around $100^3$ $\bm{k}$ points.
The rapid convergence of the latter method is attributed to the smearing effect of temperature.
